# DSTT-MRAM: Differential Spin Hall MRAM for On-chip Memories

Yusung Kim, Sri Harsha Choday, and Kaushik Roy

School of Electrical and Computer Engineering, Purdue University, West Lafayette, Indiana, 47907, USA

A new device structure for spin transfer torque based magnetic random access memory is proposed for on-chip memory applications. Our device structure exploits spin Hall effect to create a differential memory cell that exhibits fast and energy-efficient write operation. Moreover, due to inherently differential device structure, fast and reliable read operation can be performed. Our simulation study shows 10X improvement in write energy over the standard 1T1R STT-MRAM memory cell, and 1.6X faster read operation compared to single-ended sensing (as in standard 1T1R STT-MRAMs). The bit-cell characteristics are promising for high performance on-chip memory applications.

## 1) Introduction

Spin transfer torque magnetoresistive random access memory (STT-MRAM) is a promising candidate for on-chip memories due to its high-density, non-volatility, and compatibility with CMOS technology[1-3]. However, high performance memory design is quite challenging with standard 1T1R STT-MRAM. Large write current (voltage) is required for high-speed write which imposes severe stress conditions on the tunneling oxide in the magnetic tunnel junction (MTJ), leading to reliability concern such as time-dependent dielectric breakdown (TDDB)[3-6]. Consequently, the upper limit for MTJ voltage, or equivalently, the maximum achievable write-speed, is determined by the tunnel barrier reliability. One approach to address such reliability issues is to design the MTJ with lower critical switching current[1-3,5,6]. However, when switching current is lowered, the read current must also reduce in order to prevent disturb-failures (accidental write during read)[3,5,6]. The reduced read current then translates into increased sensing time, thus slowing the single-ended read operation. As a result, achieving high speed for both write and read is challenging with standard 1T1R STT-MRAM due to its limited design space.

The spin-transfer torque induced by the spin current in spin-Hall effect (SHE) is a promising methodology for manipulating the magnetization of a nanomagnet[8-14]. Since the effective spin injection efficiency can exceed 100%, the write mechanism using SHE has the potential to be fast (>1GHz) and energy efficient (<0.1pJ/bit). In this letter, we propose an STT-MRAM bit-cell that exploits the transverse spin currents in SHE to store both true and complimentary bits for the same energy required to write the true bit. This results in a memory structure that is self-referenced for a full-differential and fast read operation.

## 2) Proposed DSTT-MRAM structure

DSTT-MRAM consists of two MTJs and a spin Hall metal (SHM) which is a non-magnetic conductor with spin orbit interaction (see Fig. 1(a)). The free layers of the MTJs are in contact with the SHM and the magnetizations of the free layers ($\mathbf{m}_1$ and $\mathbf{m}_2$) represent the stored information. The pinned layer magnetization of the two MTJs is fixed in one direction.

Our proposed structure utilizes spin Hall effect to switch the magnetizations of $\mathbf{m}_1$ and $\mathbf{m}_2$ for an energy-efficient write. In the example shown in Fig. 1(b), $\mathbf{m}_1$ and $\mathbf{m}_2$ are initially oriented along the +X and –X direction, respectively, and a charge current is flowing in the SHM in +Y direction. The coupling between electron spin and orbital motion (spin-orbit coupling) in SHM deflects the –X and +X directed spins to +Z (top) and –Z (bottom) surfaces of the SHM. As a result, the accumulated spins on the top and bottom surfaces exert spin transfer torque (STT) on $\mathbf{m}_1$ and $\mathbf{m}_2$. Because spin Hall effect separates opposite spins to opposite surfaces, $\mathbf{m}_1$ and $\mathbf{m}_2$ will be subjected to equal magnitude but opposite sign of spin currents, writing both true and complementary bits at the same time. Note that the write energy used for writing both true and complementary bits is same as the energy required to write the true bit. Furthermore, in the proposed device, the write current can be increased for high-speed (<1ns write) without reliability constraint imposed by TDDB, as no tunneling current is required. The use of spin Hall effect for write provides write energy improvement over the standard 1T1R STT-MRAMs which can be explained as follows. In standard STT-MRAM, spin-polarized electrons are injected to the free-layer through tunneling and each electron exerts only one quanta of angular momentum to the free-layer. On the other hand, in spin Hall effect based write, one electron can exert STT many times by repeatedly scattering at the interface between the SHM and the free-layer[8]. As a result, the spin current injected to the free layer can be larger than the charge current, leading to an energy-efficient write.

The differential read operation in the proposed memory structure is illustrated in Fig. 1(b), where the SHM acts as the common terminal for the read current paths. In this figure, $MTJ_1$ and $MTJ_2$ are in the parallel (P) and anti-parallel (AP) states, respectively. The resistance of the MTJs is dependent on the orientation of the free layer with respect to the fixed layer; hence the difference between the two read current (voltage) is sensed to evaluate the stored bit. Unlike standard STT-MRAM, our proposed device is self-referencing and does not require a global reference cell.

The read and write operations in DSTT-MRAM are performed by applying appropriate voltages to the write word line (WWL), read word line (RWL), two bit-lines (BL/BLB), and source-line (SL). In the bit-cell shown in Fig.2, WWL and RWL control the access transistors and are asserted high during write and read operation, respectively. To write '0', BL/BLB are driven to the required write voltage ($V_{WRITE}$) and SL is driven to ground so that charge current flows from BL/BLB to SL through the SHM. To write '1', the voltage polarity of BL/BLB and SL are reversed so that the write current flows in the opposite direction. To read out the state of $MTJ_1$ and $MTJ_2$, the RWL is asserted high and read current is applied to BL/BLB while SL is at ground. The corresponding voltages on the BL and BLB are compared to determine the data stored in the MTJs.

**3) Modeling and Simulation**

The DSTT-MRAM has a different current path for read and write, hence the equivalent circuits during read and write operations are different. As shown in Fig. 3(b), the equivalent circuit during the read operation consists of $MTJ_1$ and $MTJ_2$ in series with their access transistors. The resistance of the MTJs in parallel ($R_P$) and anti-parallel ($R_{AP}$) states is obtained from the non-equilibrium Green's function (NEGF) based simulation framework[15]. The NEGF formalism uses a spin dependent effective mass Hamiltonian for electron transport simulations, which were calibrated with experimental data to reproduce the resistance characteristics of the MTJ reported in (ref. 16). Subsequently, the voltage-dependent resistances ($R_P$ and $R_{AP}$) of the MTJs were used in SPICE based circuit simulations along with a commercial 45nm transistor model to evaluate the read performance of the proposed bit-cell. A voltage-sensing scheme was used to read the data, where a read current is applied to the bit-cell and the corresponding voltages on BL and BLB are measured (see Fig. 3(b)). Note that the voltages on the bit-lines can either be $V_P$ or $V_{AP}$ corresponding to the resistance of the MTJ.

During write, the equivalent bit-cell is a resistor (resistance of SHM) in series with an access transistor as shown in Fig. 4(b). The charge current ($I_e$) flowing through the SHM is extracted from SPICE based circuit simulation and the corresponding spin current is calculated from the expression shown below[9].

$$I_S = \frac{A_{MTJ}}{A_{SHM}} \theta_{SHM} \left(1 - \text{sech}\left(\frac{t_{SHM}}{\lambda_{sf}}\right)\right) I_e \qquad (1)$$

where, $A_{MTJ}$ is the cross sectional area of the MTJ, $A_{SHM}$ is the cross sectional area of SHM, $\theta_{SHM}$ is the spin Hall angle, and $\lambda_{sf}$ is the spin flip length ($\lambda_{sf}$ =1.5nm). The term in the parenthesis accounts for the reduction in spin Hall current density as $t_{SHM}$ is reduced[14]. The spin current obtained from eq. (1) is used with the generalized Landau-Lifshitz-Gilbert (LLG) equation[17,18] to analyze the switching dynamics of $\mathbf{m_1}$ and $\mathbf{m_2}$. Note that the two free layers are separated by thin SHM. As a result, the dipolar coupling between $\mathbf{m_1}$ and $\mathbf{m_2}$ has to be taken into account. Under the macrospin approximation, the magnetization dynamics of $\mathbf{m_i}$ (i=1, 2) can be written as:

$$\frac{d\mathbf{m}_i}{dt} = -\gamma\left(\mathbf{m}_i \times \mathbf{H}_{\text{eff},i}\right) + \alpha \frac{d\mathbf{m}_i}{dt} + \frac{1}{q(M_S\Omega/\mu_B)}(\mathbf{m}_i \times \mathbf{m}_i \times I_S \mathbf{M}_i) \quad (2)$$

where, the effective field ($\mathbf{H}_{\text{eff},i}$) includes self-demagnetization field due to shape anisotropy, the uniaxial field due to magneto-crystalline anisotropy, and the interlayer dipolar field. The dipolar coupling field is given by $\mathbf{H}_{\text{Dip},i} = -N_{dip} M_s \mathbf{m}_j$ where $N_{dip}$ is effective dipolar coupling factors extracted from micromagnetic simulations[19]. Since the dipolar field inside a magnet is non-uniform, the dipolar fields from $\mathbf{m}_j$ on $\mathbf{m}_i$ were averaged over the $\mathbf{m}_i$ volume to obtain the effective dipolar coupling factors[20]. The simulations parameters are listed in Table I.

## 4) Results and Discussion

DSTT-MRAM exhibits three distinct advantages over the standard 1T1R bit-cell. First, the write current flows through SHM instead of the tunneling oxide. As a result, high write current can be supplied to achieve fast switching without any reliability concerns associated with the tunnel barrier. Second, the SHM has a much lower resistance than an MTJ; hence a smaller transistor width is needed for the write current. Finally, the write operation using SHE is more energy efficient because the spin injection efficiency can be greater than 100%. To quantitatively show the improvement of write energy compared to 1T1R STT-MRAM, we designed both DSTT-MRAM and standard 1T1R STT-MRAM with an identical target write-time of 1ns.

In the design of DSTT-MRAM, the spin Hall angle is expected to be the dominant factor in determining the spin injection efficiency. As shown in Fig. 4, for each type of SHM we find the thickness ($t_{SHM}$) at which spin injection efficiency is maximized. Note that the $t_{SHM}$ at which maximum spin current flows is different for writing '1' and '0'. When writing '0', the access transistors are driving a charge current from SL to BL/BLB, in which case the transistors have a gate-to-source voltage ($V_{GS}$) of $V_{DD}$. On the other hand, when writing '1' the direction of charge current is reversed and hence node X (see Fig. 4) acts as the source terminal of the transistor. Due to a finite voltage at node X, the $V_{GS}$ of the access transistor is less than $V_{DD}$, thus reducing its drive strength. Therefore, the observed asymmetry in the spin current is due to source degeneration of the access transistor caused by the resistance of SHM ($R_{SHM}$). To reduce the asymmetry, $t_{SHM}$ is chosen for the worst write case i.e. writing a '1'. Since spin injection efficiency approaches ~3X in DSTT-MRAM, a minimum sized access transistor (120nm) is used to achieve a 1ns write time.

In case of 1T1R cell design, the source degeneration of access transistor is due to the MTJ which has a much higher resistance than SHM. The stronger source degeneration in 1T1R cells causes much stronger write asymmetry and results in wasted write energy[21]. It is also noted that in 1T1R bit-cell design, boosted voltage (i.e. $V_{DD} = 1.2V$) and large transistor width (1.5µm) are needed to meet the write-time requirement of 1ns. As a result, in 1T1R cells, high write-speed design leads to severe stress condition on the tunnel barrier, degrading the reliability. The simulation results show that DSTT-MRAM achieves write energy of 0.077 pJ which is ~10 X smaller than that of 1T1R cell (0.823 pJ).

For reading the data stored in DSTT-MRAM, a read current is injected into BL, BLB and the corresponding voltages are sensed. For this differential cell, the sense margin ($SM$) can be written as ($V_{AP} - V_P$). Whereas in a single-ended sensing scheme (as in 1T1R bit-cell), only the true bit is stored, hence the bit-cell provides either $V_P$ or $V_{AP}$ which is compared with a reference voltage ($V_{REF}$) to determine the stored data. The value of $V_{REF}$ is typically chosen as the average of $V_P$ and $V_{AP}$. Therefore the $SM$ in single-ended read is $(V_{AP} - V_P)/2$. This two-fold increase in the sense margin due to the differential nature of DSTT-MRAM allows fast read operation. The read time is typically defined as the time taken to develop a sense margin of 50mV. Fig.3 shows the read time of single-ended and differential sensing for different thickness of MgO ($t_{MgO}$). It is shown that the optimum $t_{MgO}$ occurs at 1.2nm and differential read achieves 1.6X faster read than that of single-ended sensing. Note that in 1T1R bit-cell, $t_{MgO}$ affects both read and write performance, and hence, thinner $t_{MgO}$ may have to be chosen to improve write performance.

## 5) Conclusion

We propose a differential STT-MRAM bit-cell that utilizes spin-Hall effect, resulting in a write operation that consumes 10X less energy than standard 1T1R STT-MRAMs. Moreover, storing both true and complimentary bits comes without any write energy overhead due to the nature of spin Hall effect. The proposed DSTT-MRAM bit-cell can perform 1.6X faster read due to an inherently differential device structure. As a result, DSTT-MRAM is suitable for high performance on-chip memories.

## 6) Acknowledgements

This work was supported in part by STARnet, Semiconductor Research Corporation, and by Intel Corporation.

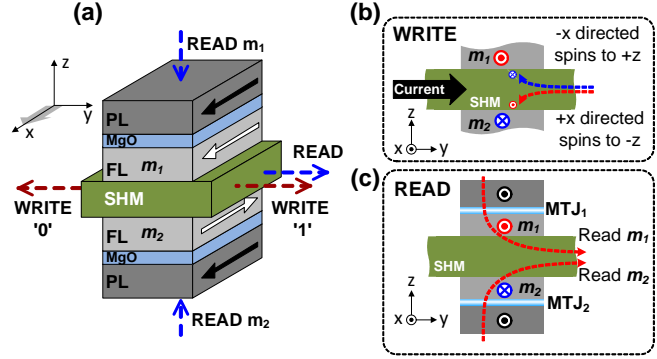

FIG. 1: (a) Proposed memory device structure and current paths for (b) write and (c) read operations.

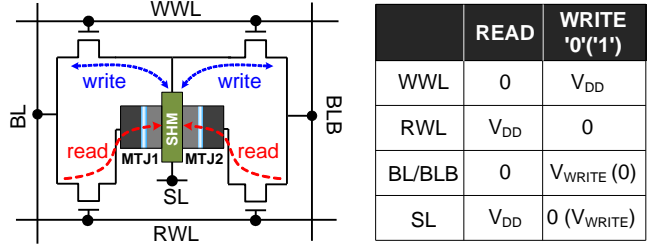

FIG. 2: A possible bit-cell structure and the biasing conditions for read and write operations.

| Simulation Parameters | |
|---|---|
| Gilbert Damping, $\alpha$ | 0.0122 |
| Saturation Magnetization, $M_S$ | 850 kA/m |
| Free layer volume, $\Omega$ | $40 \times (2.89 \times 40) \times 2$ nm$^3$ |
| Spin Hall metal dimension | $80 \times (2.89 \times 40) \times 2.8$ nm$^3$ |
| Spin Hall angle (W,Ta,Pt) | (0.3, 0.12, 0.08) |
| Spin Hall metal resistivity (W,Ta,Pt) | (200, 190, 20) $\mu\Omega\cdot$cm$^2$ |
| Self-Demagnetization factors | (0.022, 0.066, 0.911) |
| Mutual dipolar coupling factors | (0.0098, 0.030, -0.039) |
| Magneto-crystalline anisotropy, $H_{Ku2}$ | 5 mT/$\mu_0$ |
| MgO thickness, $t_{MgO}$ | 1.2nm |
| CMOS Technology / $V_{DD}$/ $V_{WRITE}$ | 45nm SOI CMOS / 1V / 0.4V |
| Access transistor width | 120nm |

Table I: Simulation parameters[9-14]

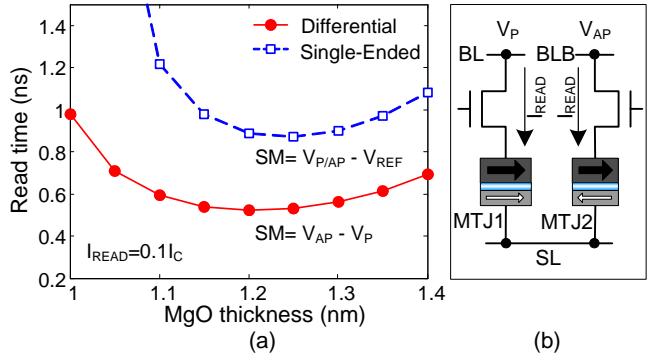

FIG 3: (a) Read time (time to develop 50mV of read signal margin) for differential and single-ended read (b) equivalent bit-cell during read

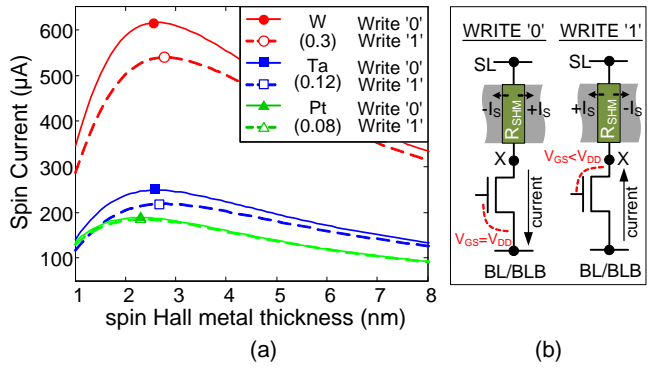

FIG 4: (a) Spin current during write operation for different spin Hall metal thickness and materials (b) equivalent bit-cell during write